\documentclass[aps,prd,twocolumn,preprintnumbers,nofootinbib,showpacs]{revtex4}
\usepackage{graphicx}
\usepackage{epsfig}
\usepackage{amssymb}
\usepackage{color}
\begin{document}

\title{Quark number susceptibilities at high temperatures}

\author{
A. Bazavov$^{\rm a}$,
H.-T. Ding$^{\rm a,b}$,
P. Hegde$^{\rm c}$,
F. Karsch$^{\rm a,c}$,
C. Miao$^{\rm d}$,
Swagato Mukherjee$^{\rm a}$,
P. Petreczky$^{\rm a}$, 
C. Schmidt$^{\rm c}$, 
A. Velytsky$^{\rm a}$
}
\affiliation{
$^{\rm a}$ Physics Department, Brookhaven National Laboratory, Upton, New York 11973, USA \\
$^{\rm b}$ Physics Department, Columbia University, New York, New York 10027, USA\\
$^{\rm c}$ Fakult\"at f\"ur Physik, Universit\"at Bielefeld, D-33615 Bielefeld, Germany\\
$^{\rm d}$ Institute of Nuclear Physics,Johannes Gutenberg-Universit\"at Mainz,
Johann-Joachim-Becher-Weg 45, D-55099 Mainz, Germany
}

\begin{abstract}
We calculate second and fourth order quark number susceptibilities 
for 2+1 flavor QCD in the high temperature region.
In our study we use two improved staggered fermion formulations,
namely the highly improved staggered quark formulation, and the so-called
p4 formulation, as well as several lattice spacings.
Second order quark number susceptibilities are
calculated using both improved staggered fermion formulations,
and we show that in the continuum limit the two formulations give consistent results.
The fourth order quark number susceptibilities are studied only using the p4 formulation and at non-zero
lattice spacings. We compare our results on quark number susceptibilities
with  recent weak coupling calculations, and find that these agree reasonably well
with the lattice calculations within the estimated uncertainties.
\end{abstract}

\maketitle

\section{Introduction}

At high temperatures, strongly interacting matter undergoes a deconfinement transition 
to a new state, where thermodynamics 
can be described in terms of quark and gluonic degrees of freedom. 
Studying the properties of this new state of matter is the subject of
a large effort in lattice QCD, see Refs. \cite{Petreczky:2012rq,Philipsen:2012nu}.
This is in part due to the fact that non-perturbative effects could be important
even at very high temperatures due to infrared problems \cite{Linde:1980ts}.
Therefore, it is important to clarify using lattice QCD calculations at
what temperatures the deconfined medium can be described as weakly interacting
quark-gluon gas. This is especially important in light of the recent experimental
findings showing that the matter created in ultra-relativistic heavy ion collisions behaves
like a strongly coupled liquid \cite{Muller:2006ee}.

Fluctuations of conserved charges are known to be sensitive probes of deconfinement
and suitable for testing the weakly (or strongly) interacting nature of the deconfined
medium. 
They are defined as the derivatives of the pressure with respect to the corresponding
chemical potentials. Fluctuations of conserved charges 
are expected to be exponentially small in the low temperature region 
where the conserved charges are carried
by massive hadrons. However, they are not suppressed at high temperatures,
where the dominant degrees of freedom are light quarks.  Therefore,
fluctuations of conserved charges are good probes of deconfinement.

In 2+1 flavor QCD there are three chemical potentials corresponding to
baryon number, electric charge and strangeness, or equivalently to 
$u,d$ and $s$ quark chemical potentials. Since at high temperatures
the dominant degrees of freedom are quarks and gluons, the quark number 
basis provides a natural way to study the fluctuations. In this paper
we study second and fourth order quark number fluctuations, also known
as quark number susceptibilities, defined as
\begin{equation}
\left . \chi_{2n}^q=\frac{\partial^{2n} (p/T^4)}{\partial (\mu_q/T)^{2n}} \right|_{\mu_q=0},~~~q=u,d,s,
~ n=1,2.
\label{def}
\end{equation}
Though in our calculations the $u$ and $d$ quark masses are degenerate,  the corresponding
chemical potential are always assumed to be different, i.e., we consider single flavor susceptibilities.
Often in weak coupling calculations the chemical potential for different degenerate quark flavors is
considered to be the same. In this case one effectively calculates the baryon number susceptibilities
(up to normalization factors) that are different from the single flavor quark number susceptibilities
defined in Eq. (\ref{def}).
In this paper we are interested in the high temperature behavior of the
quark number susceptibilities and comparison of the lattice results with weak coupling calculations.
To better control the continuum
extrapolation we used two different improved staggered quark formulations
in our calculations,
namely the so-called p4 action \cite{Karsch:2000kv} and the 
highly improved staggered quark (HISQ) action
\cite{Follana:2006rc}.

Fluctuations of conserved charges have been studied in lattice QCD for many years and 
confirm the expected temperature dependence discussed above. Second order quark number
susceptibilities have been studied by several groups 
\cite{Gavai:2002jt,Bernard:2004je,Aoki:2006br,Aoki:2009sc,Borsanyi:2010bp,
Bazavov:2010sb,Borsanyi:2011sw,Bazavov:2012jq}. 
Fourth order fluctuations have also
been studied using the so-called p4 improved staggered fermion action 
\cite{Allton:2003vx,Allton:2005gk,Cheng:2008zh}. More recently, they have been studied
using HISQ and stout actions \cite{Bazavov:2012vg,Bazavov:2013dta,Bellwied:2013cta,Borsanyi:2013hza}.

At sufficiently high temperatures
one should be able to describe quark number susceptibilities using weak coupling techniques.
Lattice QCD calculations of the quark number susceptibilities thus can provide useful tests
for the range of validity of these weak coupling approaches.
Second order quark number susceptibilities have been calculated using resummed perturbation theory
\cite{Blaizot:2001vr,Rebhan:2003fj,Haque:2010rb}
as well as the dimensionally reduced effective theory \cite{Vuorinen:2002ue,Hietanen:2008xb}, for some time.
The fourth order quark number susceptibilities have been first calculated in perturbation theory in Ref. \cite{Vuorinen:2003fs},
and very recently have also been studied using resummed perturbation theory \cite{Andersen:2012wr,Mogliacci:2013mca,Haque:2013qta,Haque:2013sja}.
In this paper, we make 
precision tests of the applicability of various weak coupling calculation techniques by performing continuum
extrapolated lattice QCD calculations of second order quark number susceptibilities at even higher
temperatures. 
The analysis of various second and fourth order susceptibilities \cite{Bazavov:2013dta} suggests
that the deconfined medium becomes weakly interacting for $T > 300$ MeV.

We extend our earlier calculations of diagonal quark number susceptibilities in 2+1 flavor
QCD to higher temperatures and compare them with perturbative calculations.
The rest of the paper is organized as follows. In Sec. II we present 
the details of the lattice simulations. Section III is dedicated to the discussion
of the cutoff effects of quark number susceptibilities in the free theory.
Our main numerical results are summarized in Sec. IV. In Sec. V we compare
our lattice results with the results of weak coupling calculations. Finally, section VI
presents our conclusions.
Some preliminary results have been presented in conference 
proceedings \cite{Petreczky:2009cr,Bazavov:2012gm}.

\section{Details of the lattice simulations}

We performed calculations in 2+1 flavor QCD using p4 and HISQ actions at
the physical value of the strange quark mass $m_s$. The gauge configurations
have been generated using the rational hybrid Monte Carlo algorithm 
\cite{Clark:2004cp}. 
The lattice spacing has been fixed using the $r_1$ and $r_0$ scales defined in terms
of the static potential
\begin{equation}
\left . r^2 \frac{dV}{dr} \right|_{r_x}=C_x\;\; ,\;\; x=0,1,
\end{equation}
where $C_0=1.65$ and $C_1=1.0$. 
The parameter $r_0$ is also known as the Sommer scale \cite{Sommer:1993ce}.
As in Ref. \cite{Bazavov:2011nk} we use the values
$r_0=0.468$~fm and $r_1=0.3106$~fm. 
The lattice spacing in units of $r_0$ and $r_1$ as function of the
bare gauge coupling was given in Ref. \cite{Cheng:2007jq} for the p4 action and
in Ref. \cite{Bazavov:2011nk} for HISQ action.
The quark number susceptibilities can be expressed in terms
of the quark matrix and the corresponding formulas were given in Refs.
\cite{Allton:2002zi,Allton:2003vx,Allton:2005gk}.
The necessary operators are evaluated using the random noise method 
(see Ref. \cite{Allton:2002zi} for details).

We calculated second order quark number susceptibilities 
using the HISQ action on lattices with temporal extent
$N_{\tau}=4,~6,~8,~10,~12$ and $16$, and aspect ratio $N_{\sigma}/N_{\tau}=4$,
with $N_\sigma$ denoting the spatial extent of the lattice.
For light quark masses, we used $m_l=m_s/20$, corresponding to a pion
mass of $160$ MeV in the continuum limit  \cite{Bazavov:2011nk}.
Our calculations covered a wide temperature range from $200$ to about $950$ MeV
(1400 MeV for $N_{\tau}=4$). We used 20 random noise vectors to evaluate the operators
needed for quark number susceptibilities for $N_{\tau}=4$ and $6$ lattices, $50$ random noise
vectors for $N_{\tau}=8$ lattices, $100$ random noise vectors for $N_{\tau}=10$ lattices, and
$200$ random noise vectors for $N_{\tau}=12$. We accumulated 3000 to 8000 molecular
dynamics trajectories of unit length for each temperature value.

For the p4 action the calculations have been performed 
on $32^3 \times N_{\tau}$ lattices with temporal extent $N_{\tau}=6,~8$ 
and $12$. The light quark mass in this calculation was $m_l=m_s/10$, 
corresponding to a pion mass of $220$ MeV in the continuum limit \cite{Cheng:2007jq}.
For $N_{\tau}=6$ and $8$ lattices we accumulated between 10,000 and 25,000 trajectories,
while for $N_{\tau}=12$ we accumulated between 3000 to 10,000 trajectories.
The length
of the molecular dynamic trajectory here was $0.5$.
On $N_{\tau}=6$ and $N_{\tau}=8$ lattices, we calculated quark number susceptibilities both
for the strange and light quarks.    
We typically used 196 random vectors in these calculations. 
For $N_{\tau}=12$ in the light quark sector we only calculated second order quark number 
susceptibilities. Here we used 96 random vectors to evaluate the necessary operators.
In the strange quark sector, both second and fourth order quark number susceptibilities
have been evaluated and 196 random vectors have been used.   
Note that in the studied temperature range, the slightly larger than physical 
light quark masses do not have any significant effect \cite{Cheng:2009zi,Borsanyi:2010cj}.
In fact, at these high temperatures, the light quark masses in temperature units are negligibly
small and can be considered to be zero.

\section{Quark number susceptibilities in the free theory}

\begin{figure}
\includegraphics[width=8cm]{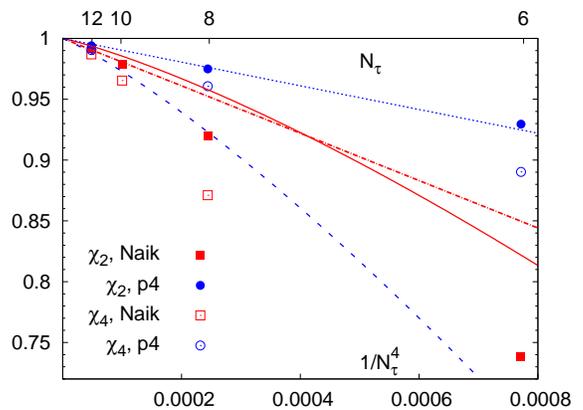}
\caption{Second (filled symbols) and fourth order (open symbols) 
quark number susceptibilities in the free
theory as functions of $N_{\tau}^{-4}$ normalized by the corresponding 
continuum Stefan-Boltzmann values. 
Also shown in the figure are the free theory results expanded in $1/N_{\tau}$ up
to ${\cal O}(N_{\tau}^6)$ \cite{Hegde:2008nx}. These are shown 
as dotted, solid, dashed, and dashed-dotted lines corresponding to $\chi_2$ and p4 action, 
$\chi_2$ and Naik action, $\chi_4$ and p4 action, and $\chi_4$ and Naik action,  
respectively. }
\label{fig:free}
\end{figure}
As a first step toward understanding the cutoff dependence of
quark number susceptibilities and making continuum extrapolations
at high temperature, let us review their cutoff dependence  in
the non-interacting massless theory. Since we consider the massless
case, the index $q$ in the quark number susceptibilities will be omitted. 
The cutoff dependence of quark number
susceptibilities has been studied in Refs. \cite{Allton:2003vx,Bernard:2004je,Hegde:2008nx}.
The p4 action and the HISQ action both contain a three-link term in the Dirac
operator and completely eliminate ${\cal O}(1/N_{\tau}^2)$ discretization
errors at tree level. In the case of the p4 action bended three-link terms
are used, while the HISQ action has a straight three-link path, 
known as the Naik term \cite{Naik:1986bn}.
The different types of the three-link terms not only eliminate the ${\cal O}(1/N_{\tau}^2)$
discretization effects but also have the same corrections at order $1/N_{\tau}^4$
\cite{Hegde:2008nx,Heller:1999xz}. The
differences between the Naik action and the p4 action appear at higher orders of $1/N_{\tau}$ 
\cite{Hegde:2008nx,Heller:1999xz}. 
In Ref. \cite{Hegde:2008nx} the $N_{\tau}$ dependence of quark number
susceptibilities was calculated analytically up to ${\cal O}(1/N_{\tau}^6)$
\begin{eqnarray}
&
\displaystyle
\chi_2=1+a_{24}\frac{\pi^4}{N_{\tau}^4}+a_{26}\frac{\pi^6}{N_{\tau}^6},\nonumber\\[2mm]
&
\displaystyle
\chi_4=\frac{6}{\pi^2}\left( 1+a_{44}\frac{\pi^4}{N_{\tau}^4}+
a_{46}\frac{\pi^6}{N_{\tau}^6} \right),\nonumber\\
&
\displaystyle
a_{24}=-\frac{93}{70},~a_{44}=-\frac{21}{10},\nonumber\\
&
\displaystyle
a_{26}=-\frac{381}{70},~~a_{46}=\frac{62}{945},~~{\rm Naik}\nonumber\\
&
\displaystyle
a_{26}=\frac{127}{3150},~~a_{46}=-\frac{62}{7},~~{\rm p4}.
\end{eqnarray}
Note, however, that numerically higher order terms in $1/N_{\tau}$ may
be important for $N_{\tau} \le 8$.
As one can see from the above equation, there is a large coefficient 
in front of the $1/N_{\tau}^6$ term in $\chi_2$
for the Naik action and a small one for the p4 action. For $\chi_4$ the situation is just the opposite.
In Fig. \ref{fig:free}, we show the complete results on the cutoff dependence of $\chi_2$ 
and $\chi_4$ and
contrast it with the analytic results.  
For second order quark number susceptibility and p4 action, the analytic result 
expanded up to order $1/N_{\tau}^6$ describes the complete
cutoff dependence fairly accurately. In all other cases the truncated analytic results give only
a qualitative description of the complete result. The complete result indicates a quite small
cutoff dependence for both $\chi_2$ and $\chi_4$ in the case of the p4 action.
It should be stressed that the free theory results in general
cannot be used for a quantitative description of the cutoff dependence
of the lattice data as higher order perturbative corrections are important and could be significant 
\cite{Heller:1999xz}. In particular, there are cutoff effects proportional to $g^{2n}/N_{\tau}^2$
both for the p4 and HISQ actions.
Yet the analysis of the cutoff dependence in the free theory provides 
an important starting point for the discussion of the cutoff effects
in the high temperature lattice calculations.

\section{Results}
In this section we present our numerical results for $\chi_2^q$ and $\chi_4^q$
and discuss the cutoff effects in quark number susceptibilities, 
as well as the details of the continuum extrapolations.
Our main result is summarized in Fig. \ref{fig:chi2cont}. Readers not interested
in technical details are advised to skip the following text till the end of
Sec. IV.B where this result is discussed.

\subsection{Numerical results on second order quark number susceptibilities}
\begin{figure*}
\includegraphics[width=8.6cm]{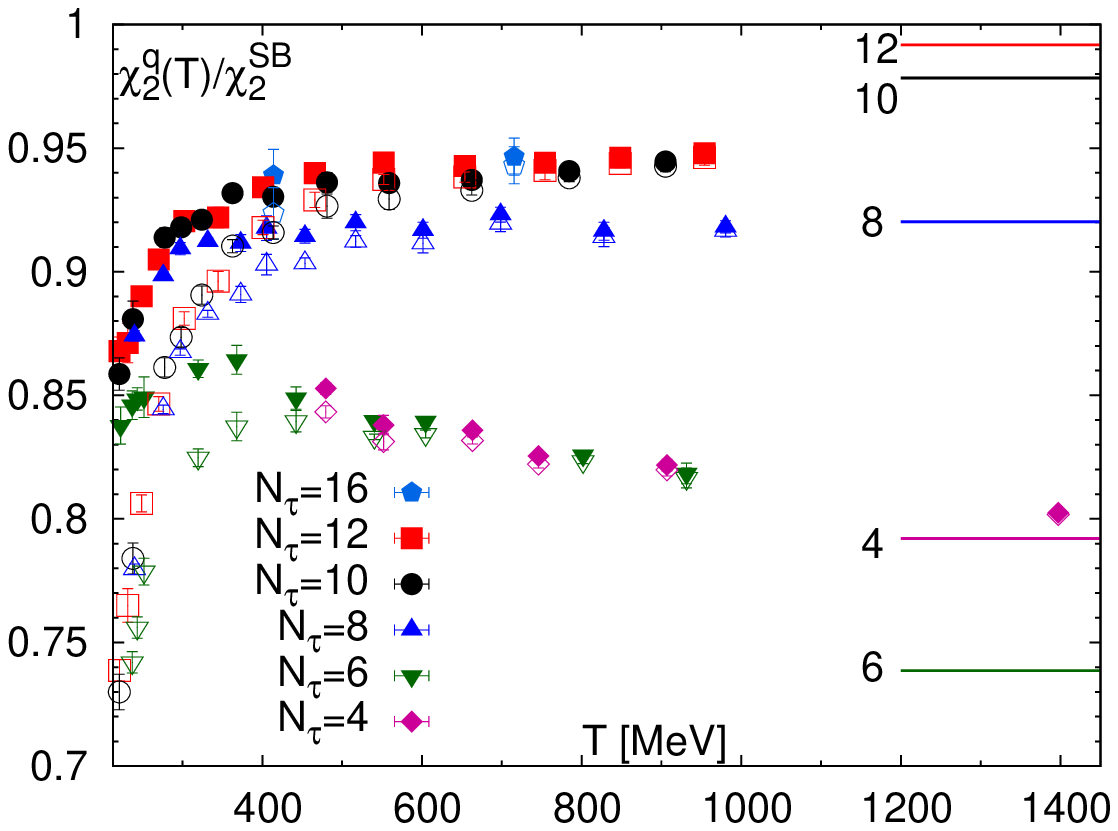}
\includegraphics[width=8.6cm]{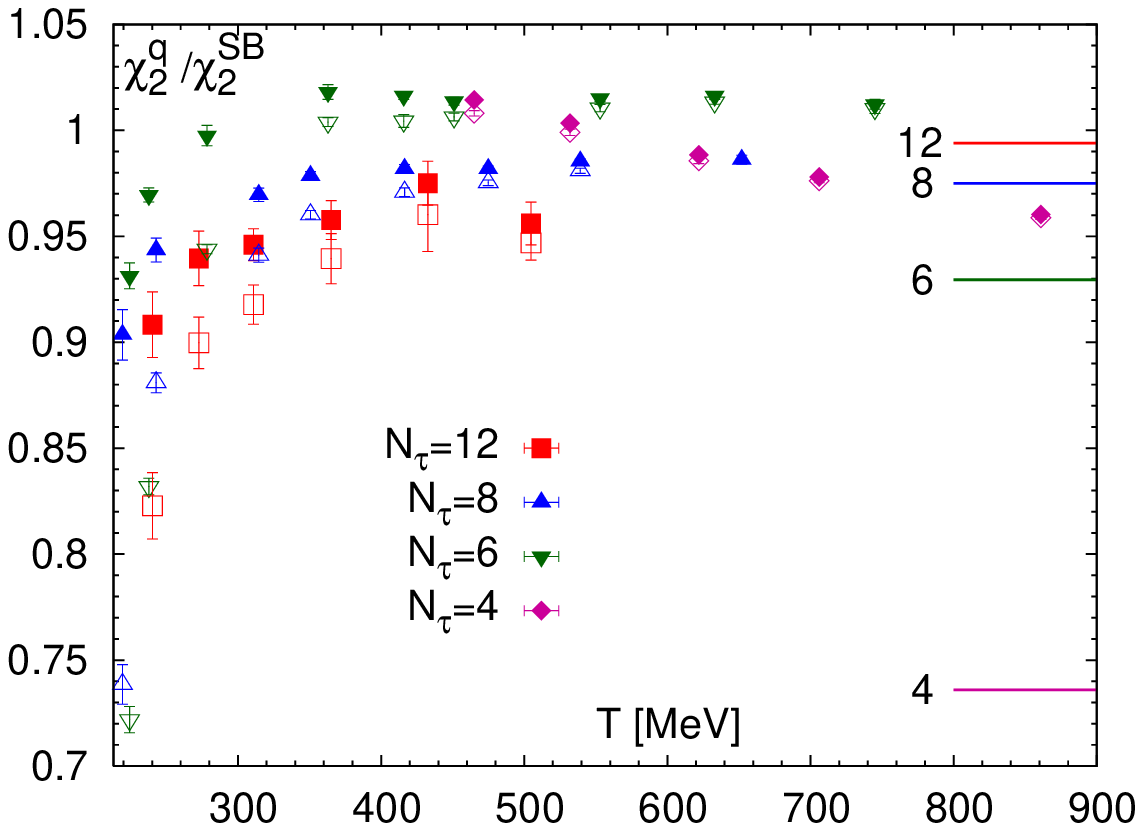}
\caption{Second order quark number susceptibilities calculated on different lattices 
with the HISQ action (left) and the
p4 action (right) as functions of the temperatures. The filled symbols
correspond to light quark number susceptibilities, while open symbols correspond
to strange quark number susceptibilities. The horizontal lines refer to the values
of the quark number susceptibilities in the free theory corresponding to different
temporal extent $N_{\tau}$. The free theory result for $N_{\tau}=4$ and the p4 action 
has been shifted upwards by 0.2 for better visibility. Note that $\chi_2^{SB}=1$. }
\label{fig:chi2all}
\end{figure*}
We start the discussion of our numerical results with the second order 
light and strange quark number susceptibilities.
In Fig. \ref{fig:chi2all}, we show the numerical results for the HISQ and p4 actions in
the high temperature region, $T>200$ MeV. In the continuum limit the quark number 
susceptibilities approach unity at very high temperatures.
The strange quark number susceptibilities approach the high temperature limit slower
than the light quark number susceptibilities. The difference between the light and strange
quark number susceptibilities becomes small above temperatures of $400$ MeV 
and negligible
for $T>600$ MeV. The difference between the light and strange quark number
susceptibilities can be understood 
in terms of the strange quark mass for $T>250$ MeV, but at lower temperatures
it is significantly larger than the expected suppression due to the relatively 
large strange quark mass, $m_s \simeq 90$ MeV, 
see the discussion in Ref. \cite{Cheng:2008zh}.
The cutoff effects are relatively small for $T<300$ MeV but
become significant above that temperature. 

In the case of the HISQ action
there is a qualitative change in the behavior of the cutoff effects for 
$T>300$ MeV; i.e., the
ordering  of quark number susceptibilities calculated at different $N_{\tau}$ 
qualitatively starts to follow expectations based on the systematics seen
for the free theory. 
The continuum limit seems to be approached from below.
The free theory result, shown as horizontal
lines in Fig. \ref{fig:chi2all} shows larger cutoff dependence 
than the numerical data. This is expected to some extent; from the analysis
of the high temperature limit of pure gauge theories \cite{Boyd:1996bx},
it is known that cutoff effects in the interacting
theory typically are about a factor 2 smaller than in the free theory. 
Interestingly enough though, at
the highest temperature the $N_{\tau}=4$ data point is very close to the 
lattice ideal gas value. 

For the p4 action, the pattern of the cutoff dependence observed in the free
theory is not seen in the interacting case for the entire temperature range
explored by us.
As discussed in Sec. III, for improved actions the quark number 
susceptibilities in the free theory approach
the continuum limit from below. Our p4 lattice data, on the other hand, seem to approach
the continuum limit from above. This implies that cutoff effects proportional to $\alpha_s$
and higher orders in the coupling constant are significant, contrary to the case of
the HISQ action. The main difference between the p4 and HISQ action that could 
be responsible for the difference in the cutoff behavior is the use of smeared gauge
fields in the latter. The use of smeared gauge fields is known to reduce cutoff effects in higher order
perturbative corrections in lattice calculations \cite{Hasenfratz:2002vv}.

\subsection{Continuum extrapolation of second order quark number susceptibilities}

Using the lattice results for quark number susceptibilities at different
$N_{\tau}$, we perform continuum extrapolations. First, for each
$N_{\tau}$ we interpolate the lattice results using smooth splines.
The errors on the interpolated values were estimated using the bootstrap
method. We used the R package for this analysis \cite{Rpackage}.
Using the interpolations, we obtain the values of the 
quark number susceptibilities at the same set of temperatures.
Finally, we perform continuum extrapolations at this set of temperatures.
The $N_{\tau}=4$ data have not been used in the continuum extrapolations.

In the case of the HISQ action we consider the temperature interval 
from $225$ to $950$ MeV for each $N_{\tau}$, 
with the step of $25$ MeV for $T \le 400$ MeV, 
and the step $50$ MeV for larger temperatures.
Our extrapolations for HISQ are motivated by the leading
order $N_{\tau}$ dependence of quark number susceptibilities in
the free theory. Namely, we performed continuum extrapolations
using the following form:
\begin{equation}
\chi_2^q(N_{\tau})=a+b/N_{\tau}^4+c/N_{\tau}^6.
\label{fitform}
\end{equation}
We also performed extrapolations
using the simpler $a+b/N_{\tau}^4$ form and data for $N_{\tau}\ge 8$ only.
The two fits gave identical results within the estimated errors.
Furthermore, we used the complete tree-level result for the $N_{\tau}$ 
dependence
of the quark number susceptibility to perform the continuum extrapolation;
i.e., we fitted the data for each temperature with $a+c \cdot (\chi_2^{q,free}(N_{\tau})-\chi_2^{SB})$.
This gives extrapolated values for $\chi_2^q$ that are systematically lower
than the above fits, but still agree within errors. 
For the coefficient $c$, we typically find values around $0.6$; i.e. the
cutoff effects in the interacting theory are $40\%$ smaller than
in the free field limit.
Since beyond tree level there are also discretization errors proportional
to $1/N_{\tau}^2$, we also performed extrapolations using 
$a+c \cdot (\chi_2^{q,free}(N_{\tau})-\chi_2^{SB})+d/N_{\tau}^2$. These extrapolations
give results that agree within errors with the extrapolations obtained
using the Ansatz $a+b/N_{\tau}^4+c/N_{\tau}^6$, though they are systematically 
higher. The coefficient $d$ turns out to be negative, and at the highest
temperatures it is compatible with zero. Thus, it is possible that 
the $1/N_{\tau}^2$ term  just mimics cutoff effects proportional to 
$1/N_{\tau}^n,~~n\ge 4$ at higher order in the weak coupling expansion.
It turns out that the differences between the mean values of $\chi_2^q$ obtained
by using the above two fits and the fit that uses Eq. (\ref{fitform}) are
smaller or equal to the statistical errors of the three-parameter fit
given by Eq. (\ref{fitform}). In other words, the systematic errors
estimated as the differences of the above three fits are smaller 
or of the same size as the statistical errors of that fit. Therefore, we use 
the extrapolation based on Eq. (\ref{fitform}) and its statistical errors
as our final continuum result for HISQ. 
 
For the p4 action the continuum limit is approached from above 
contrary to the free field expectations. This implies that the
dominant cutoff effects come from higher orders in the weak coupling expansion
and scale like $1/N_{\tau}^2$. Therefore, we performed the continuum
extrapolation using the simple constant plus $1/N_{\tau}^2$ form.
We also tried to add  an $1/N_{\tau}^4$ term in the fit when doing
the continuum extrapolations. It turns out that the inclusion of
such a term did not change the result within the errors. Moreover,
the coefficient of the $1/N_{\tau}^4$ term was 4-10 times smaller
than in the free theory. This confirms our assumption that the dominant
cutoff effects in the case of the p4 action go like $1/N_{\tau}^2$.
\begin{figure}
\includegraphics[width=9cm]{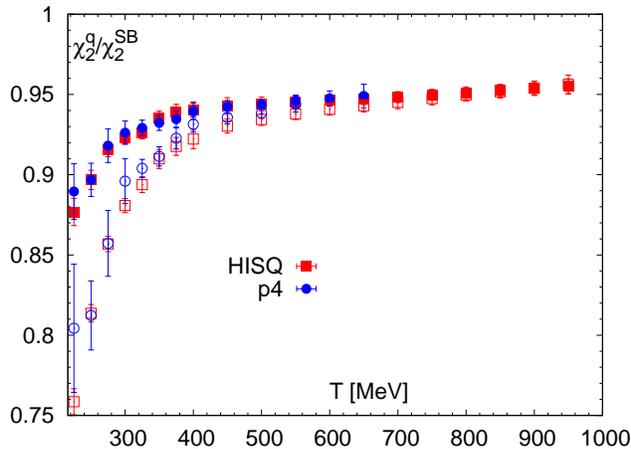}
\caption{Second order light (filled symbols) and strange (open symbols)
quark number susceptibilities as functions of the temperature in the
continuum limit. Note that $\chi_2^{SB}=1$.}
\label{fig:chi2cont}
\end{figure}

\begin{figure*}
\includegraphics[width=5.8cm]{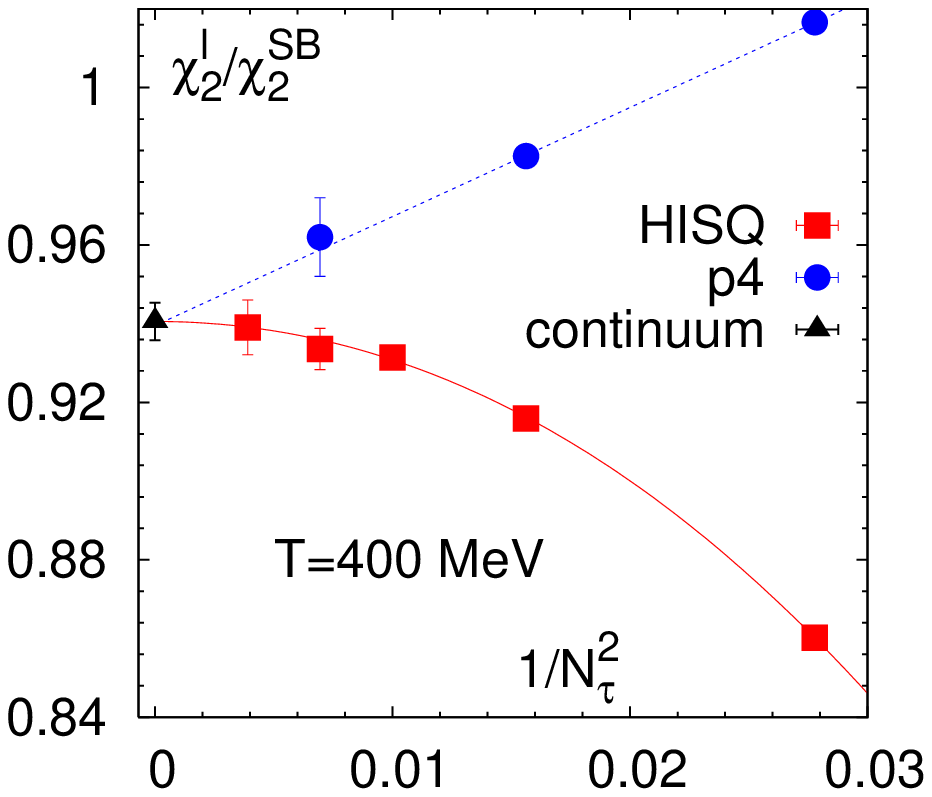}
\includegraphics[width=5.8cm]{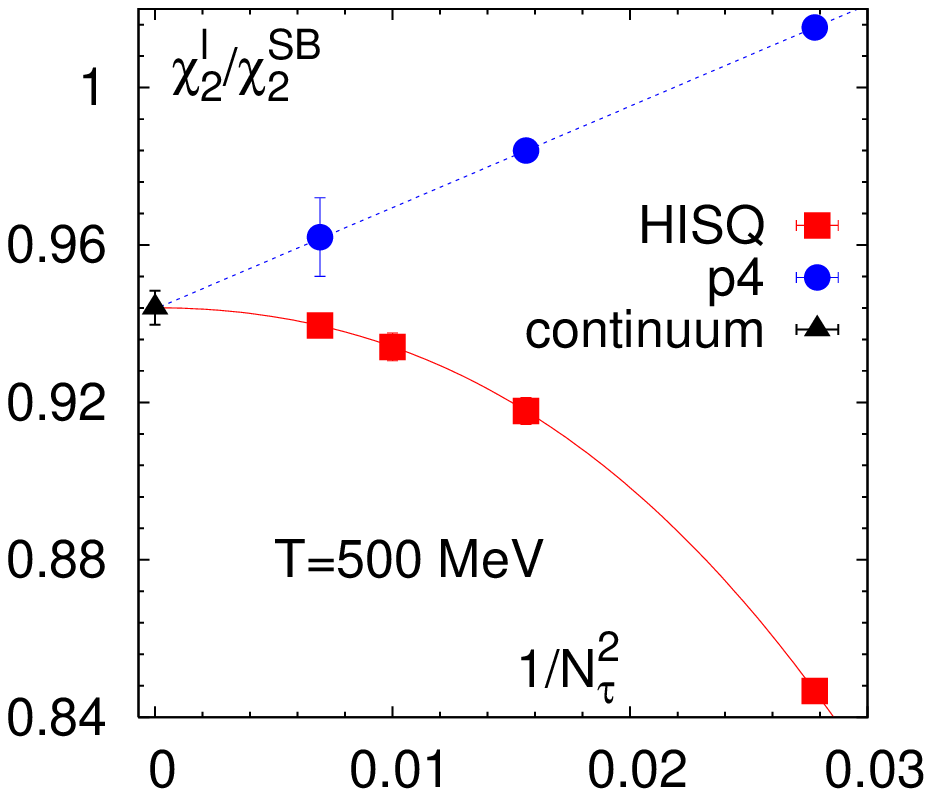}
\includegraphics[width=5.8cm]{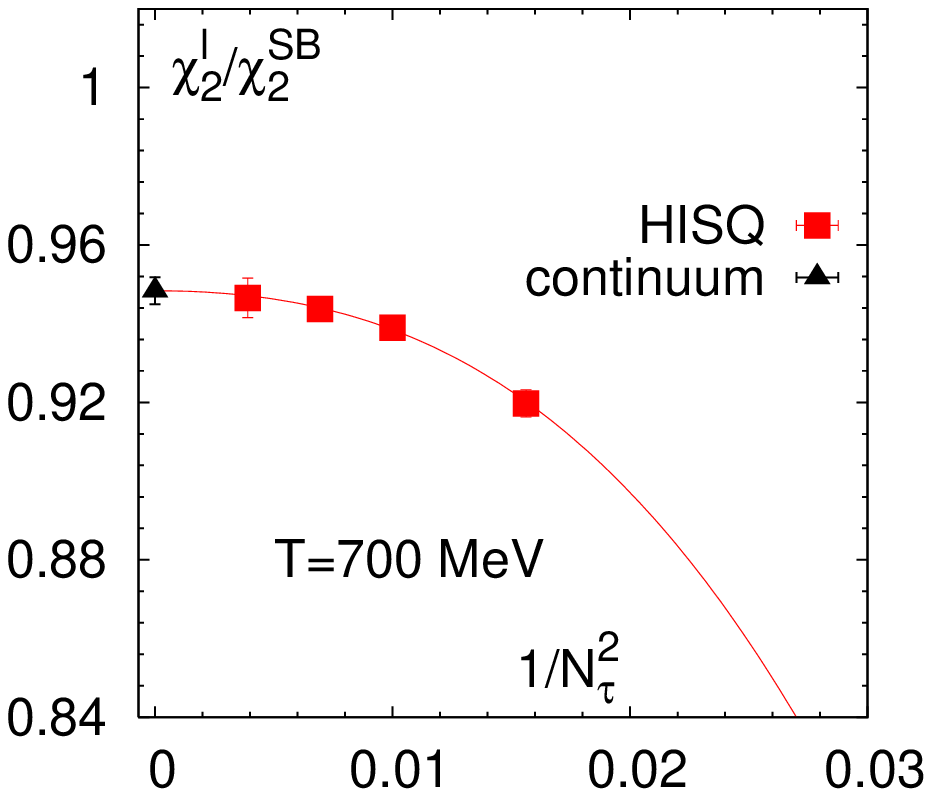}
\caption{The $N_{\tau}$ dependence of $\chi_2^l$ obtained with HISQ and p4 actions
for $T=400$ MeV (left), $500$ MeV (middle) and $700$ MeV (right, HISQ only). The lines correspond
to the fit of $N_{\tau}$ dependence based on Eq. (\ref{fitform}) and $1/N_{\tau}^2$ form
for HISQ and p4 action respectively. 
The triangles correspond to the continuum value obtained using the HISQ action.
The $N_{\tau}=16$ data have not been included
in the fit.}
\label{fig:chi2Nt}
\end{figure*}

The continuum extrapolated quark number susceptibilities are shown
in Fig. \ref{fig:chi2cont} for both p4 and HISQ actions. Overall, the p4 results
and HISQ results agree quite well.
Note that the agreement between the p4 results and HISQ results
is particularly good for $T>400$ MeV. 
This is remarkable in view of the different nature of the cutoff effects for the
HISQ and p4 actions. To better illustrate this point, in Fig. \ref{fig:chi2Nt} we
show the $N_{\tau}$ dependence of $\chi_2^l$ for the p4 and HISQ action together with the continuum extrapolations
based on the $1/N_{\tau}^2$ form and Eq. (\ref{fitform}), respectively, 
for two temperatures $T=400$ MeV and $T=500$ MeV. 
In the figure we also show the $N_{\tau}$ dependence of the HISQ results for $T=700$ MeV together
with the fit based on Eq. (\ref{fitform}).
Note that the $N_{\tau}=16$ HISQ data
point was not included in the continuum extrapolations but happens to lie on the
fitted curve. 

For the p4 action the cutoff effects at order $1/N_{\tau}^4$
and higher seem to  be very small, and the dominant cutoff effects are proportional to
$1/N_{\tau}^2$ with a positive coefficient. For the HISQ action, there is no
indication for such a term in the data, and if it is put in the Ansatz
for the extrapolations,
the corresponding coefficient turns out to be negative.
The cutoff dependence, thus, is well described
by the free theory  modified by a multiplicative factor.
The continuum extrapolated values for $\chi_2^q$ obtained
using HISQ action are given in Table \ref{tab}.
Remarkably, the continuum extrapolated $\chi_2^q$ results
at high temperatures deviate from the massless ideal gas limit only by $5\%$. 
We also have compared our results  with recent continuum extrapolated results 
obtained by using the stout action \cite{Borsanyi:2011sw}. For $\chi_2^s$, 
our result agrees with
the stout results within errors up to $350$ MeV. For $350$ MeV $<T<400$ MeV
the stout results are lower by $(1.2-1.5)$ standard deviations. For $\chi_2^l$
our results agree with the stout results only up to $260$ MeV. Above
that temperature the stout results are lower than ours by 2 standard
deviations. 
\begin{table}
\begin{tabular}{|c|c|c|}
\hline
$T$ [MeV] & $\chi_2^l$ & $\chi_2^s$ \\
\hline
225 & 0.8768(85) & 0.7585(81) \\
250 & 0.8968(60) & 0.8137(55) \\
275 & 0.9156(44) & 0.8568(48) \\
300 & 0.9230(38) & 0.8807(44) \\
325 & 0.9263(40) & 0.8939(49) \\
350 & 0.9352(44) & 0.9098(59) \\
375 & 0.9393(46) & 0.9176(56) \\
400 & 0.9406(47) & 0.9222(60) \\
450 & 0.9428(52) & 0.9303(44) \\
500 & 0.9440(43) & 0.9344(37) \\
550 & 0.9452(39) & 0.9379(35) \\
600 & 0.9462(36) & 0.9407(36) \\
650 & 0.9472(35) & 0.9430(38) \\
700 & 0.9484(34) & 0.9451(40) \\
750 & 0.9497(35) & 0.9474(39) \\
800 & 0.9511(35) & 0.9497(38) \\
850 & 0.9526(37) & 0.9516(38) \\
900 & 0.9539(41) & 0.9538(45) \\
950 & 0.9551(47) & 0.9563(58) \\
\hline
\end{tabular}

\caption{Continuum extrapolated values for the second order light and strange quark number
susceptibilities obtained using HISQ action.}
\label{tab}
\end{table}

\subsection{Fourth order quark number susceptibilities}
\begin{figure}
\includegraphics[width=8.8cm]{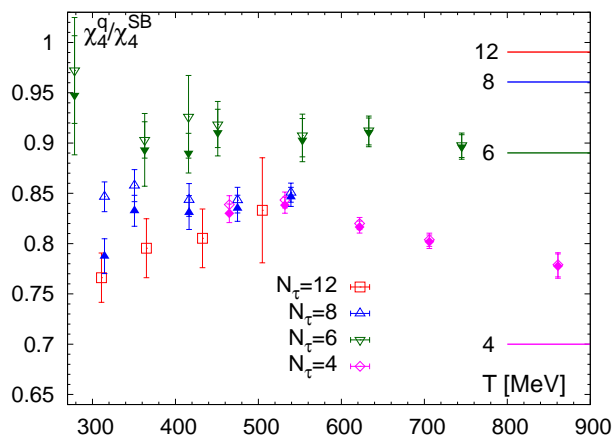}
\caption{Fourth order quark number susceptibilities calculated
with the p4 action and normalized by the corresponding Stefan-Boltzmann value.
The horizontal lines correspond to the free field
value of $\chi_4^q/\chi_4^{SB}$ in the massless limit. 
The free field result for $N_{\tau}=4$ has been shifted upwards
by $0.44$ for better visibility.
The open symbols correspond
to the strange quark, while the filled symbols correspond to the light
quarks. Note that $\chi_4^{SB}=6/\pi^2$. }
\label{fig:chi4}
\end{figure}
As mentioned already in Sec. II, for the p4 action we also calculated
the strange and light fourth order quark number susceptibilities for $N_{\tau}=6$ and $8$.
The calculation of the fourth order light quark number susceptibility is more
demanding than the calculation of the corresponding strange quark number susceptibility.
Therefore, for $N_{\tau}=12$ we calculated the fourth order quark number susceptibility
only in the strange quark sector. 
Our numerical results for $\chi_4^q$ normalized by the corresponding massless 
ideal gas (SB) value are shown in Fig. \ref{fig:chi4}. 
The cutoff dependence
of $\chi_4^q$ is qualitatively the same as for $\chi_2^q$; namely, the continuum
limit is approached from above, contrary to the free theory results shown in 
Fig. \ref{fig:chi4} as  horizontal lines.
As discussed above, the difference between the light and strange quark number
susceptibilities is expected to be small for $T>400$ MeV and our numerical data
clearly show this. In fact, with the exception
of the $N_{\tau}=8$ data point at the lowest temperature,
 the difference between the light and strange fourth order
quark number susceptibilities is of the same order or smaller than the statistical
errors. For $\chi_4^q$ the deviations from the ideal gas value seem to
be significantly larger than for $\chi_2^q$ at temperatures $350$ MeV $<T<500$ MeV, and increase with increasing 
$N_{\tau}$.

Given the large statistical errors of the $N_{\tau}=12$ lattice data, it is
at present difficult to perform a reliable continuum 
extrapolation for $\chi_4^s$. However,
the $N_{\tau}$ dependence of $\chi_4^s$ for $350$ MeV $<T<500$ MeV is compatible with the $1/N_{\tau}^2$
behavior of the cutoff effects, and such an extrapolation would result in
the value of $\chi_4^s/\chi_4^{q,SB}$ around $0.77$; i.e., the deviations from the massless ideal
gas limit for $\chi_4$ could be almost 4 times larger than for the second order quark
number susceptibilities.  Looking at the data shown in Fig. \ref{fig:chi4}, it is possible
that the ordering of $N_{\tau}=8$ and $N_{\tau}=12$ data will change for $T>550$ MeV,
becoming qualitatively compatible with the free theory expectations. 
Therefore, it would be very important to extend the
lattice calculations of fourth order susceptibility to higher temperatures.

\section{Comparison of the lattice and the weak coupling results}
\begin{figure*}
\includegraphics[width=8.5cm]{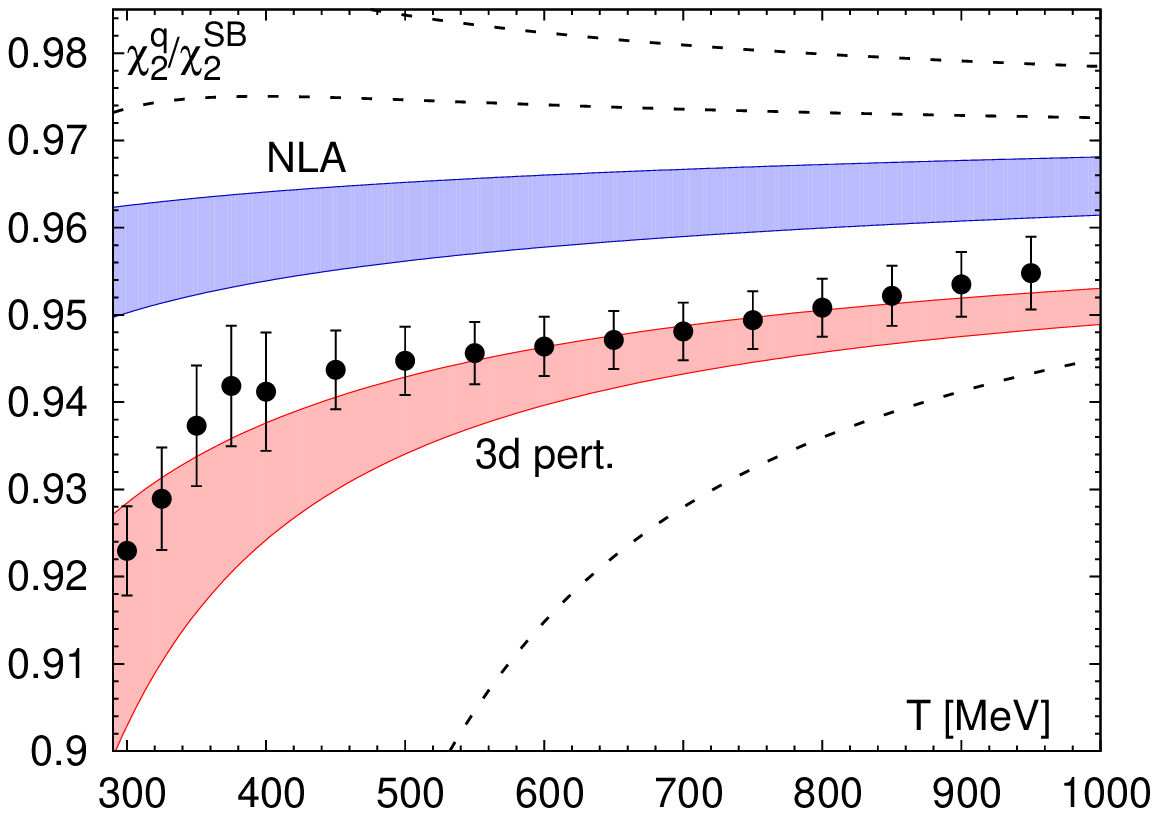}
\includegraphics[width=8.5cm]{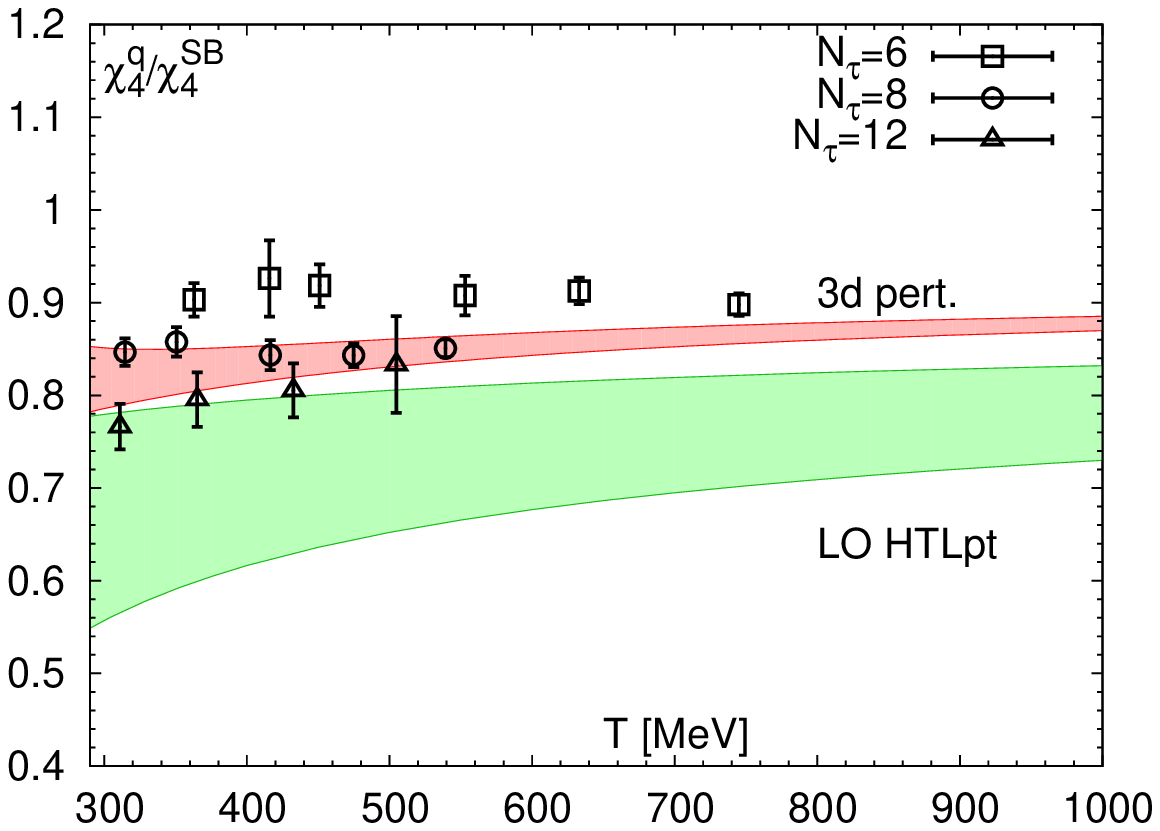}  
\caption{The second order quark number susceptibility (left)
and fourth order quark number susceptibility (right) calculated
on the lattice and in resummed perturbation theory.
For the second order quark number susceptibilities we show
the results obtained in dimensionally reduced 3d effective theory \cite{Mogliacci:2013mca},
the results of NLA calculations \cite{Rebhan:2003fj}, and of three-loop HTLpt (dashed lines) \cite{Haque:2013sja},
corresponding to different renormalization scales $\mu=\pi T,~2 \pi T$ and $4 \pi T$ (bottom to top).
For the fourth order quark number susceptibilities we show the results obtained in
reduced 3d effective theory and the one-loop HTLpt \cite{Mogliacci:2013mca}.
The ideal massless quark gas susceptibilities are  $\chi_2^{SB}=1$ and $\chi_4^{SB}=6/\pi^2$.}
\label{fig:pt}
\end{figure*}

Let us finally compare our lattice findings with results 
from weak coupling calculations. As already discussed in
Sec. I weak coupling results have existed for the second order quark number
susceptibility for some time. Recently, resummed perturbative calculations for
the fourth order quark number fluctuations also became available 
\cite{Andersen:2012wr,Mogliacci:2013mca,Haque:2013qta,Haque:2013sja}. 
In these studies, also the second order quark number susceptibilities 
have been considered. 

In Fig. \ref{fig:pt}, we
show the comparison of our continuum extrapolated data for $\chi_2^l$
with the perturbative calculations in the dimensionally reduced effective
field theory \cite{Mogliacci:2013mca} 
\footnote{
We compare here with the updated version of the
3d resummed perturbative calculation shown in
Ref. \cite{Mogliacci:2013mca}. We thank S. Mogliacci and
A. Vuorinen for bringing to our a problem with an
earlier version of the perturbative calculation
for $\chi_2^l$  and sending us
the corrected version of it prior to publication.
}.
The scale in the dimensionally reduced calculations was fixed using
the criteria of the fastest apparent convergence for the gauge coupling of the dimensionally 
reduced effective theory \cite{Kajantie:1997tt}
and was varied around
this optimal value by a factor of 2.
The width of the band shown in Fig. \ref{fig:pt} correspond to this variation of
the scale as well as to the uncertainty in the value of the gauge coupling \cite{Mogliacci:2013mca}.
As mentioned in Sec. I, in the case of more than one quark flavor, one
has to distinguish between the case where different quark flavors couple to different or the 
same chemical potential. In the latter case there are contributions from off diagonal 
quark number susceptibilities $\chi_{11}^{ij},~i\neq j, i=u,d,s$ which are absent in the
single flavor quark number susceptibilities defined in Eq. (\ref{def}). 

The resummed perturbative calculations in Refs. \cite{Rebhan:2003fj,Haque:2013qta,Haque:2013sja}
consider the case of three degenerate massless quark flavors with equal chemical potential. 
Therefore, the corresponding results cannot be directly compared with our lattice calculations
of $\chi_2^l$. However, we also estimated the second order off-diagonal quark number susceptibilities
 $\chi_{11}^{ij}$ and find that their contribution is about $1\%$ or less in the temperature range
considered here. Furthermore, since 
$\chi_{11}^{ij}$ has very large statistical errors and is only marginally different from zero,
its inclusion would not change the comparison of our lattice data and the results of the resummed
perturbative calculations. Therefore, in Fig. \ref{fig:pt} we also compare our results with
resummed perturbative calculations in so-called next-to-leading log approximation (NLA) \cite{Rebhan:2003fj} shown
as the blue band and to the three-loop hard thermal loop perturbation theory (HTLpt) \cite{Haque:2013sja}.
The width of the blue band
reflects the scale uncertainties of the NLA calculations and corresponds to
the variation of the renormalization scale from $\mu=\pi T$ to $\mu=4 \pi T$.
The dashed lines shown in Fig. \ref{fig:pt} correspond to the 3-loop HTLpt results
for the renormalization scale  $\mu=\pi T, ~2 \pi T$ and $4 \pi T$.

As one can see from  Fig. \ref{fig:pt}, the resummed perturbative calculation based on
the dimensionally reduced effective theory reproduces the lattice results reasonably well,
while the calculation performed in NLA approximation is above the lattice data by a few
percent. The three-loop HTLpt results also agree with the lattice results, given their quite large
scale uncertainty. Note, however, that the one-loop HTLpt results of Refs. \cite{Haque:2013qta,Mogliacci:2013mca}
are below the lattice data. Furthermore, the calculations of Ref. \cite{Haque:2013qta} consider
the case of three degenerate quarks with equal chemical potential, while Ref. \cite{Mogliacci:2013mca}
considers single flavor quark number susceptibilities. However, both calculations yield very similar
results. This further justifies the comparison of our lattice results 
with the results of Refs. \cite{Rebhan:2003fj,Haque:2013sja}.

The comparison of the lattice and weak coupling results for the 
fourth order quark number susceptibility 
normalized to the Stefan-Boltzmann limit
is also shown in Fig. \ref{fig:pt}.
Unlike for the second order quark number susceptibilities the contribution
from the  off-diagonal terms is significant.
Here we only show the comparison of the lattice results for $\chi_4^s$ with 
resummed calculations within the effective 3d theory  and LO HTLpt calculations 
from Ref. \cite{Mogliacci:2013mca} that considers single flavor quark number susceptibilities.
The width of the bands again correspond to the scale
uncertainty of the perturbative calculations.
For the calculations within the dimensionally reduced effective theory, the 
uncertainty band was estimated in the same manner as for the second order
susceptibility (see above).
The uncertainty of the LO HTLpt calculations 
of Ref. \cite{Mogliacci:2013mca} comes from the scale uncertainty as well as the  uncertainty of the gauge coupling. 
Using the lattice data for $\chi_4^s$ for the comparison with the perturbative results 
is justified as the effects of the non-vanishing strange quark mass are smaller
than current errors in lattice calculations. Within errors
the $N_{\tau}=8$ and $12$ lattice results for $\chi_4^s$ are compatible
with the perturbative calculations. However, as already
discussed in the previous section, for temperatures below $400$ MeV
there is a clear tendency for $\chi_4^s$ to
decrease with increasing $N_{\tau}$. 
Thus, in the continuum limit $\chi_4^s$
may turn out to be below the above-mentioned perturbative results, at least for $T<400$ MeV.
If we assume a $1/N_{\tau}^2$ behavior of the cutoff effects for $\chi_4^s$ (which
seems to be correct for $\chi_2^s$) the continuum limit would be
below the perturbative result. Of course, to see if this is indeed the
case, calculations with the HISQ action for the fourth order susceptibility will
be required. Let us mention that the preliminary continuum estimate
of the fourth order quark number susceptibility obtained with the stout action is
also below the resummed 3d perturbative result \cite{Borsanyi:2012rr}.

The fourth order baryon number susceptibility has been calculated in 3-loop
HTLpt \cite{Haque:2013sja}, and the corresponding results show agreement with  the available lattice data for $T<400$ MeV.
As discussed above, the fourth order  baryon number susceptibility differs significantly from the 
fourth order quark number susceptibility by off-diagonal contributions. Therefore, to further
validate the perturbative calculations, it will be important to also calculate off-diagonal susceptibilities,
which we plan to do in the near future.

\section{Conclusions}
We have calculated second and fourth order quark number susceptibilities
for $T>200$ MeV in lattice QCD using two different improved staggered
fermion formulations: the p4 and HISQ actions. 
We performed continuum extrapolations for the second order quark number
susceptibilities.While the cutoff dependence 
of the quark
number susceptibilities is quite different for the HISQ and p4 actions, we obtain
consistent results in the continuum limit. 
This makes us confident that the
continuum extrapolations are under control. The detailed study of the cutoff effects
in the quark number susceptibilities at high temperatures provides valuable
information for analyzing the cutoff dependence of the pressure
and other thermodynamic quantities, where the numerical calculations are
much more involved due to the need of proper vacuum subtractions.
In particular, we find indications that the cutoff effects in the temperature
interval $400$ MeV $<T< 950$ MeV are $40\%$ smaller than in the free theory.

Lattice calculations of quark number susceptibilities provide stringent
constraints on the applicability of resummed perturbative calculations at high temperature.
We performed a detailed comparison with the available perturbative results
and find  agreement with them given the uncertainties 
of the latter. To further constrain the reliability of the perturbative results,
it will be important to perform continuum extrapolations for the fourth order
quark number susceptibilities as well as to study off-diagonal quark number 
susceptibilities.

\noindent
{\bf Acknowledgments:} 
This work was supported by the U.S. Department of Energy under
Contract No. DE-AC02-98CH10886. The
numerical computations have been carried out on the
QCDOC computer of the RIKEN-BNL research center, on the QCDOC computer and the PC clusters of the USQCD
Collaboration in FNAL, the BlueGene/L computer at the New York Center for Computational
Sciences (NYCCS), and in NERSC. P.P. is also grateful to the Department of Atomic Physics 
of E\"otv\"os University, for its hospitality, where this manuscript was partly finalized and
to N. Haque, S. Mogliacci, M. Mustafa, A. Rebhan, M. Strickland, and A. Vuorinen for correspondence.

\bibliographystyle{h-physrev.bst}
\bibliography{HotQCDa}

\end{document}